\documentclass[aip,rsi,reprint]{revtex4-1}

\usepackage{hyperref}
\usepackage{float}
\usepackage{graphicx} 
\usepackage{dcolumn} 
\usepackage{bm} 
\usepackage[utf8]{inputenc}
\usepackage[T1]{fontenc}
\usepackage{mathptmx}
\usepackage{placeins} 

\begin{document}

\title{Ultra-low noise, bi-polar, programmable current sources}

\author{M. S. Mrozowski}
\email{marcin.mrozowski@strath.ac.uk}
\author{I. C. Chalmers}%
\author{S. J. Ingleby}
\author{P. F. Griffin}
\author{E. Riis}
\affiliation{SUPA, Department of Physics, Experimental Quantum Optics and Photonics Group, University of Strathclyde, John Anderson Building, 107 Rottenrow East, Glasgow G4 0NG, United Kingdom}

\date{\today}

\begin{abstract}
	We present the design process and implementation of fully open-source, ultra-low noise programmable current source systems in two configurations. Although originally designed as coil drivers for Optically Pumped Magnetometers (OPMs), the device specifications make them potentially useful in a range of applications. The devices feature a bi-directional current range of $\pm$~10~mA and $\pm$~250~mA respectively on three independent channels with 16-bit resolution. Both devices feature narrow 1/f noise bandwidth of 1~Hz, enabling magnetic field manipulation for high-performance OPMs. They exhibit low noise of 146.3~pA/$\sqrt{\mathrm{Hz}}$ and 4114~pA/$\sqrt{\mathrm{Hz}}$ which translates to 14.57~ppb/$\sqrt{\mathrm{Hz}}$ and 16.46~ppb/$\sqrt{\mathrm{Hz}}$ noise relative to full scale.
\end{abstract}

\maketitle

\section{Introduction}
Precise current control is a vital tool in many scientific applications. Examples include: driving laser diodes\cite{doi:10.1063/1.1593783}, characterisation of semiconductor devices\cite{doi:10.1063/1.4794734}, high impedance tomography and spectroscopy \cite{Goren2018}, and magnetic field manipulation\cite{doi:10.1063/5.0027848,doi:10.1063/5.0002964,doi:10.1063/1.5087957}.
Each of these applications differs in requirements, but all of them benefit from stability, low noise, and accuracy.

The work presented in this paper is informed by the development of atomic optical magnetometers, where the development is focused on different types of Optically Pumped Magnetometers (OPMs) ranging from unshielded RF to zero-field Spin-exchange relaxation free (SERF). OPMs are used in a wide range of applications, such as geosurveying\cite{doi:10.1190/1.2133784}, magnetoencephalography\cite{Boto2018} (MEG) or magnetocardiography (MCG)\cite{Yang2021}. In these applications, current sources act as coil drivers to provide an adjustable source of a magnetic field.

Static magnetic field coils are used for either nulling external magnetic fields, such as Earth's magnetic field, presenting a bias field in a direction of interest,  or for testing magnetometers by providing accurate and stable magnetic fields \cite{PhysRevApplied.14.064067}.

Compensation coils are used in zero field magnetometers such as SERF magnetometers \cite{Kominis2003,Yang2021} which operate in a low field of $\pm$~20~nT \cite{PhysRevLett.89.130801}. For this reason, Earth's magnetic field needs to be suppressed. The primary geomagnetic background field can be reduced with the use of $\mu$-metal shielding that is internally degaussed \cite{doi:10.1063/1.3491215}, or by placement in magnetically shielded rooms (MSR). There is, however, some magnetic residue that remains, which needs to be reduced further by the means of active compensation. This active compensation must be stable and feature a low noise contribution over the bandwidth of the magnetometer so as to not limit the sensitivity of the device. 

SERF magnetometers have high sensitivity (in the order of fT/$\sqrt{\mathrm{Hz}}$ \cite{Kominis2003}) and narrow low frequency bandwidth, which makes them ideally suited for applications such as MEG. Low-frequency magnetic noise performance is crucial, as the signals produced by the brain reside in the 0.1~-~40~Hz range, with most sensors for these applications operating between 0.1~-~150~Hz \cite{Boto2018}. It is important that the 1/f noise contribution is as narrow as possible and that the overall wideband noise performance is not a limiting factor for the magnetometer \cite{Colombo2016}. SERF sensors for MEG operate in an MSR with a typical residual field compensation requirement of 50~nT \cite{HILL2020116995}. The magnetic signals arising from the brain have a typical magnitude of 10~-~100~fT \cite{HARI2012386}. Combining the highest field present, and the lowest signal to be detected gives us a dynamic range in the order of ppb. In MEG applications, compensation magnetic fields are small; typically in the order of 1-10s of mA.
To achieve the ppb requirement, the current noise performance needs to be in the range of pA/$\sqrt{\mathrm{Hz}}$ across the bandwidth of interest.

Total field sensors, such as ones used for geosurveying, operate in the Earth's magnetic field which has a typical value of 50~$\mu$T \cite{doi:10.1063/5.0002964}. These sensors are not as sensitive as zero-field sensors but are able to operate over a much larger range of magnetic fields. The sensitivity of a typical total field OPM is on the order of <~1~pT$\sqrt{Hz}$ \cite{Belfi:07}. The dynamic range of this sensor operating in the background magnetic field of Earth would thus be <~20~ppb. The typical currents required for compensating fields in the total field sensor are in the order of 10-100s of mA due to the increased field required to compensate earth's field. To satisfy the ppb requirement, current noise performance in the range of nA/$\sqrt{\mathrm{Hz}}$ is required. In these applications bandwidth is usually small (< 100 Hz), as the current provided is used for the generation of static fields.

In this paper, we present two bipolar current source devices that target different maximum output ranges at the expense of noise performance. The low current driver (LCD) is capable of producing up to$\pm$~10~mA with a noise floor of 146.3~pA/$\sqrt{\mathrm{Hz}}$. The high current driver (HCD) is capable of producing up to $\pm$~250~mA with a noise floor of 4114~pA/$\sqrt{\mathrm{Hz}}$. Both devices are capable of independently driving three channels at 16-bit resolution and are controlled via USB. They have a common form factor measuring 160~x~100~mm and fit into a standard enclosure which provides extra shielding. One device costs approximately £200 to make on a single unit basis.

\section{System design}
\begin{figure*}
	\includegraphics[width=1\textwidth]{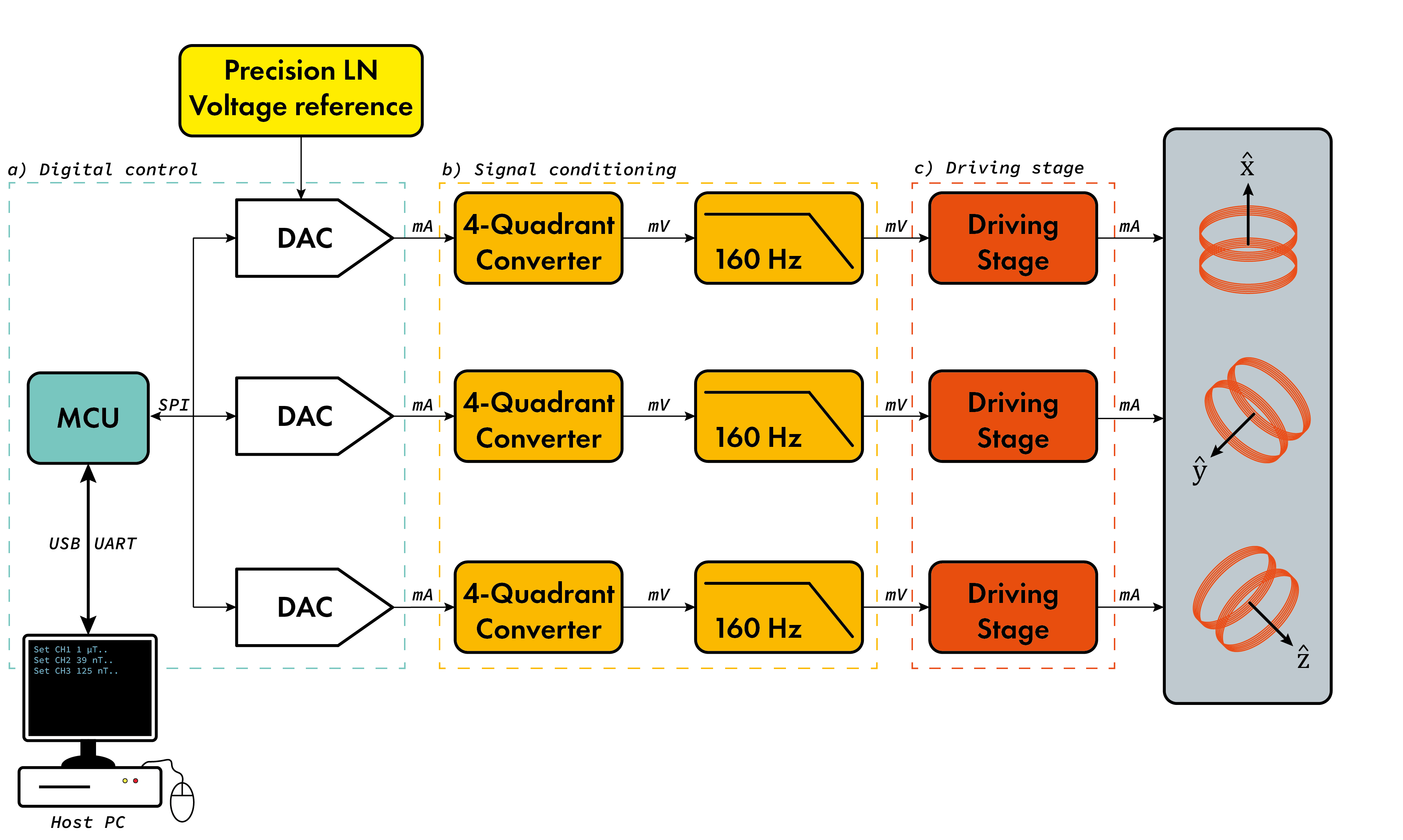}
	\caption{Simplified architecture of the coil driver system. The design is split into three sections: (a) digital control, (b) signal conditioning, (c) and driving stage. The design is realised on a four-layer printed circuit board (PCB) with the signal conditioning and driving stages protected by EMI/RFI shielding. The design files, such as the schematic, board layout and bill of materials are available on GitHub\cite{github_current_source}}
	\label{fig:VV_diagram}
\end{figure*}

The simplified architecture of the current driver is shown in Fig.~\ref{fig:VV_diagram}. The design can be broken into four sections: the external interface;  digital control (Fig.~\ref{fig:VV_diagram}a);  signal conditioning (Fig.~\ref{fig:VV_diagram}b); and output driving (Fig.~\ref{fig:VV_diagram}c). The LCD and HCD only differ in how they implement the driving stage.

\subsection{Interface}

The device can be powered from either a lab bench power supply $\pm$10-16~V or an equivalent battery source. The incoming power is regulated with a complimentary pair of TPS7A4700\cite{TI:TPS7A4700} and TPS7A33\cite{TI:TPS7A33} (Texas Instruments) low dropout regulators. These feature ultra-low noise and high power supply rejection ratio over our bandwidth of interest ensuring low noise operation without having to rely on expensive power supplies to achieve the best performance.

Each channel output is fed into a corresponding BNC connector and a single RJ45 interfacing coils to the driver. RJ45 connectors are widely used in our group, as they offer an inexpensive source of shielded twisted pairs.

\subsection{Digital control}

The current source is controlled by the ATmega4809\cite{Microchip:ATMega4809} (Microchip) present on an Arduino Nano Every board. The Arduino ecosystem was chosen for its accessibility and ease of programming for those unfamiliar with programming embedded systems. Basic firmware was written to cover general use cases. The device receives commands through a USB serial connection and control the appropriate DAC to set the output current for each channel.

The output current can be described by the transfer function in Eq.~(\ref{eq:1}), where $I_{max}$ is the highest unipolar current that the source can provide, $Max_{count}$ is the full-scale count value of the DAC (16-bit~=~65535), and $Set_{count}$ is the current count value of the DAC that has been set by the user.
\begin{eqnarray}
	I_o = \frac{2 \times I_{max}}{Max_{count}}\times(Set_{count}-\frac{Max_{count}}{2}) \label{eq:1}
\end{eqnarray}

For example, in the basic firmware provided, sending the command "\emph{<!chan 2 45547>}" would set an LCD configured for $I_{max}=10~mA$ to output +3.9~mA on channel 2

The Arduino platform makes it easy to customise the firmware for specific use cases, such as arbitrary waveform generation or triggered outputs.

\subsection{Signal Conditioning}

The primary signal used to control the current source is generated by a DAC. As the foundational signal, it is important to minimise noise at the DAC output as it will be amplified by the later conditioning stages. For this design the 16-bit DAC, DAC8814\cite{TI:DAC8814} (Texas Instruments) was selected. This is a four-channel, current multiplying DAC. The current multiplying architecture was chosen as it features excellent linearity, fast settling, and low glitch energy and allows for the output to be conditioned externally \cite{KESTER2005539}. This approach allows for a higher level of customisation of the conditioning stage, which can be tailored to the user application. By changing the transimpedance amplifier in the conditioning stage, the designer can select amplifiers that favour precision, low noise or high speed. For example, the device can be optimised for low frequency applications (such as sweeping the static field in a SERF magnetometer) or for high frequency applications such as oscillating test fields for RF magnetometers).

The noise performance of the DAC8814 is highly dependent on the noise of its reference voltage reference. Voltage references are often the noise-defining component in analogue systems, due to shot noise present in the Zener diodes that make up the reference \cite{Low_noise_electronic_system_design}. For this reason, the DAC8814 is driven by a 2.5~V precision reference LTC6655LN\cite{ADI:LTC6655LN} (Analog Devices). This voltage reference features an output filter that drastically reduces the impact of 1/f noise as well as wideband noise.

The output of the DAC8814 is a unipolar current. The conversion to bipolar output is done using a 4-quadrant converter circuit. The converter is composed of a transimpedance amplifier (TIA) followed by an inverting summing amplifier that combines the voltage reference with the output of the TIA in a 1:2 ratio. This converts the unipolar output to a bi-polar one. It is based around an OPA2210\cite{TI:OPA2210} (Texas Instruments) which is a very low 1/f noise part with excellent DC performance.

The final signal conditioning stage is a noise suppression filter that can be optimised to the desired application. For example, our reference implementation for a MEG SERF sensor implements a low pass filter (LPF) with a cutoff frequency of 160 Hz to match the expected MEG signal bandwidth, and bandwidth of the magnetometer\cite{doi:10.1063/1.5091007}. This filter is implemented as a 2nd-order Butterworth LPF in the Sallen-Key configuration. The Sallen-Key configuration was selected as it allows for inherently higher gain accuracy and stability in comparison to the multiple feedback architecture, which requires component matching for unity gain application. Sallen-Key configuration instead relies on the op-amp parameters rather than the tolerance and stability of the passive components that form the filter \cite{TI_SK}.

The signal conditioning stage and driving stages are shielded with an EMI shield on the PCB to minimise EMI induced noise on the output current.

\subsection{Driving stages}
Customisation of the digital control and signal conditioning stages requires only a change in component values. The need to provide a variety of maximum output currents requires different circuits entirely, and hence multiple different designs are necessary. Their configuration depends on the maximum current that is required and the relative noise performance that is to be achieved. We designed two driving stages tailored to specific applications: a Low Current Driver (LCD) design targeted at SERF applications and a High Current Driver (HCD) design targeted at total-field applications.

\subsubsection{Low Current Driver (LCD)}
The LCD is used for generating small currents not exceeding 25~mA, which are used in our SERF system for small bias field (0.5~$\mu$T) cancellation through the use of Helmholtz coils. This configuration aims to provide high impedance over the frequency range of interest as well as stability.
The driving stage is based around a Howland Current Pump (HCP) \cite{lightning_empiricist}. This circuit can be realised with a single amplifier and four resistors that are close in tolerance. The choice of amplifier and resistors is crucial to ensure the performance of the device \cite{TI_HCP}. The standard HCP architecture from Ref.~\onlinecite {TI_HCP} was used for the design.

The HCP was implemented using an OPA2210. This is the same part used in the signal conditioning stage, which helps reduce BOM size. The key features that make it suitable for the HCP are its architecture and the driving stage, which can source and sink up to 60~mA. It features a very low offset of 5~$\mu$V and ultra-low noise contribution in both input voltage and current. Its architecture is based around bipolar transistors, thus its 1/f noise is much smaller than that of CMOS amplifiers \cite{TI_CMOS_JFET_BI_comp}, resulting in lower overall noise in the bandwidth of interest (DC - 160~Hz). The device was also selected based on its very high common-mode rejection ratio (CMRR), which is in excess of 132 dB. This is crucial for HCPs to maintain their high impedance across a wide frequency range. This allows their use in applications requiring the generation of AC magnetic fields in excess of 10s of kHz\cite{TI_HCP}, such as generating test fields for coils for RF magnetometers.
To meet the requirement for well matched resistors, thin-film resistors with a tolerance of 0.1\% and a temperature coefficient of 25~ppm/K were selected. Such resistors offer a good balance between performance and cost.

The maximum practical current that the LCD can be configured to provide is \textasciitilde50~mA. After this point the resistors forming the HCP begin to heat-up unevenly. The resulting decrease in CMMR degrades the noise floor and accuracy of the driver. The minimum practical current that the device can be configured to provide at full range is \textasciitilde25~$\mu$A. In this configuration, the amplifier is still capable of resolving 16-bits of resolution provided by the control circuitry. Values smaller than this would be consumed by the input bias current which would lead to non-monotonicity of the output. It is important to note that at these currents, connections to the coils may become an issue as stray current pickup can become a significant source of error.

\subsubsection{High Current Driver (HCD)}
\begin{figure}[tb]
	\centering
	\includegraphics[width=0.482\textwidth]{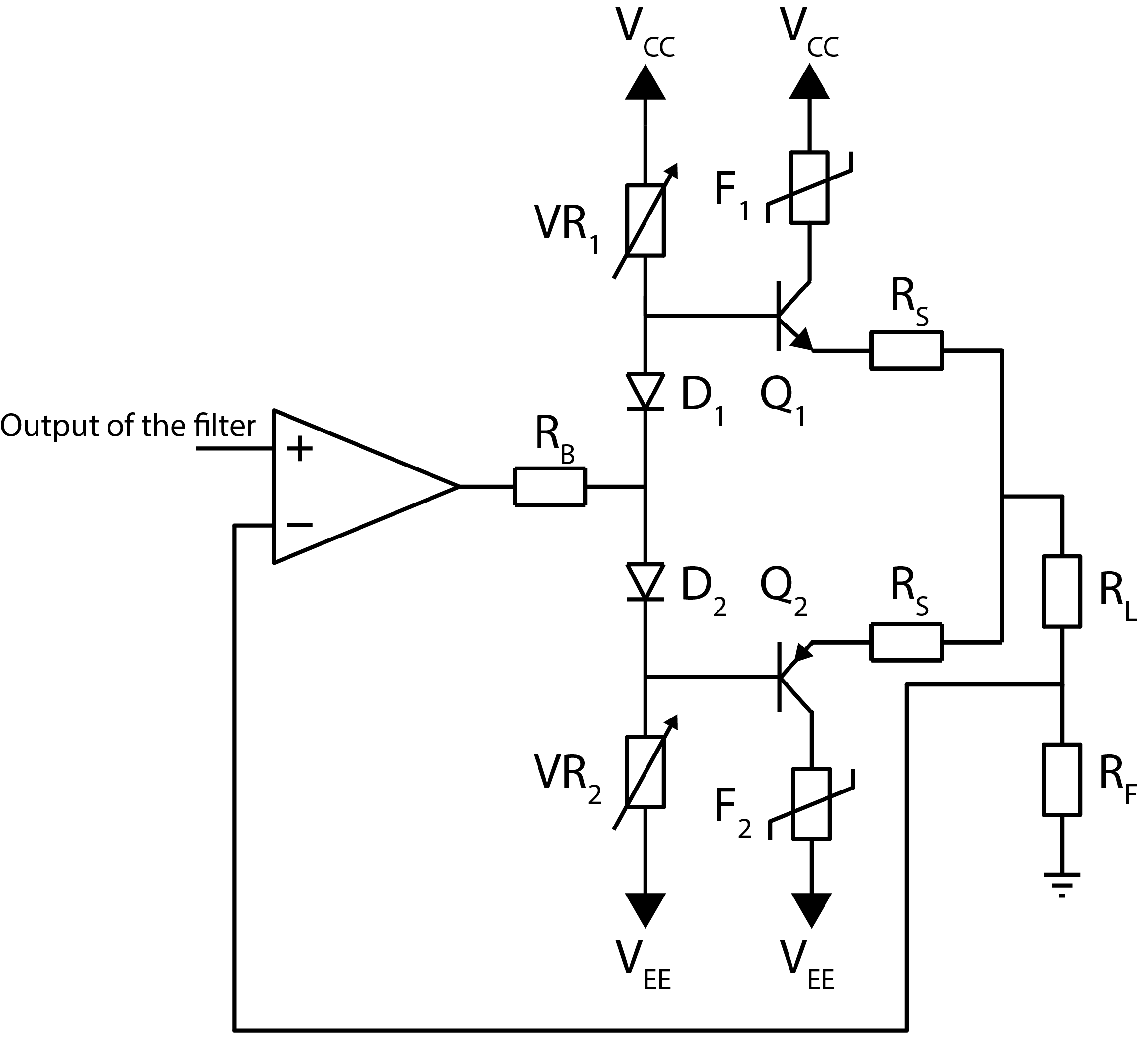}
	\caption{Simplified schematic of a single output channel of the HCD. The design uses a class AB stage driven by an OPA2210 op-amp. Complete schematics and design files can be found on GitHub\cite{github_current_source}}
	\label{fig:HCCD_arch}
\end{figure}

\begin{figure*}
	\includegraphics[width=1\textwidth]{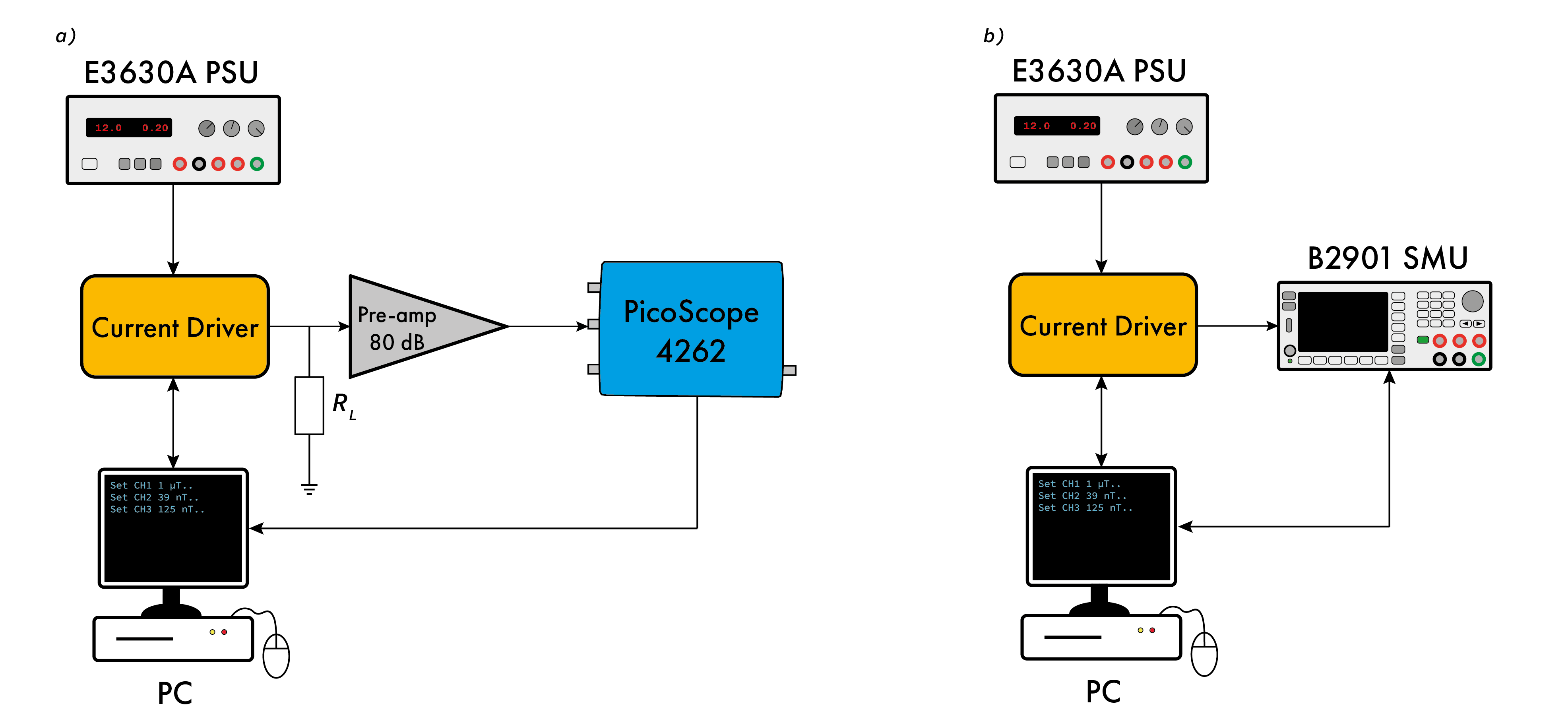}
	\caption{a). Block diagram of the test setup used for noise measurement. The current driver was powered up from a $\pm$12 V PSU (E3630A). The current driver was connected to the laptop through a galvanically-isolated USB for communication. The outputs of the LCD were 300~$\Omega$ terminated with 0.1~$\%$ tolerant, low-temperature-coefficient resistors. For the HCD no termination was used as the current sense resistor acts as a load. The signal was further amplified by the oscilloscope and captured. b) Block diagram of the test setup used for stability and accuracy measurements. The setup is identical to the "noise measurement setup" except that the SMU is directly used as the load.}
	\label{fig:Test_setup}
\end{figure*}

Many applications, such as those with constrained coil geometries, require a higher peak current than the LCD is capable of providing. Components and circuits that can provide higher peak currents usually have a non-linear trade off in their noise performance.

The HCP output stage is unable to provide such output currents without losing efficiency, and potentially accuracy due to the self-heating of the control stage. The "improved" HCP stage could be implemented as the efficiency is much greater, however, it also requires an op-amp that can deliver 250 mA to the load. Amplifiers like these exist but they do not match the noise performance of the likes of precision amplifiers such as the OPA2210. Another disadvantage is the fact that the drive circuitry is thermally coupled to the control loop. This in turn negatively affects the accuracy as the components begin to heat up.

The solution was to utilise the existing amplifier (that was previously used as a HCP) and turn it into the control driver for a class~AB amplifier stage, which delivers the final output current. The circuit is presented in Fig.~\ref{fig:HCCD_arch}. The amplifier is implemented using a complementary pair of NPN/PNP (2SC5566/2SA2013)\cite{ONSEMI:2SA2013/2SC5566} transistors ($Q_1, Q_2$). These transistors offer good power dissipation of up to 3.5~W, high current gain and low saturation voltage, meaning that they can be driven more easily from lower voltages. The complementary pair is controlled by the OPA2210. By putting the complementary pair in the feedback loop of the OPA2210, high accuracy can be maintained while increasing the output current capability. The pair is not matched hence multi-turn trim-pots ($VR_1, VR_2$) are used for adjustment of the bias current for each arm.

The thermal run-away condition, common in class AB stages\cite{ONSEMI_TRA}, is mitigated with the use of a small heat-sink that thermally bonds the complementary pair and the series bias diodes ($D_1, D_2$), keeping them at the same temperature. To further improve thermal run-away resistance, the design is equipped with polyfuses at its collectors ($F_1, F_2$) and small 0.47~$\Omega$ series resistors ($R_s$) from each emitter in the path of the load. This configuration allows for high current generation while maintaining the high accuracy and noise performance found in the preceding stages. The class AB stage is also thermally decoupled from the OPA2210 as it is placed further away, minimising temperature effects on the control circuitry. The feedback resistor $R_F$, which determines the output current, is capable of dissipating 3~W and features a temperature coefficient of 20~ppm/K and a value of 10~$\Omega$ (for 250~mA output). Although the resistor will never dissipate more than 625~mW (for the 250~mA version) the extra clearance allows it to remain at a cooler temperature, which minimises temperature effects on accuracy. By using a single feedback resistor, the maximum output current can be easily changed by swapping a single component. Note that the output is not directly referenced to ground but rather "floats" on top of the feedback resistor $R_F$. This prevents the output from being probed with a single-ended probe that is referenced to ground. 

The maximum practical current that can be set on the HCD is \textasciitilde250~mA, which is limited by the transistor thermal performance. Higher currents can be achieved by the use of polyfuses with higher rated current and replacement of the output stage transistors. The minimum practical current that the device can be configured to is \textasciitilde50~mA. The device can be configured to output lower full-scale currents, however at that point it would be more beneficial to use the LCD which features lower noise and better accuracy.

\section{Device testing}
The LCD and HCD were tested to measure their performance in three key areas: noise, stability and accuracy. For each test, measurements were taken in a temperature-controlled lab, maintaining a constant 21\textdegree C. All of the instruments were left on for at least an hour before any measurements were taken to reach thermal equilibrium and to ensure the best accuracy and stability.

\subsection{Noise test}

The noise spectrum of the current sources was measured to assess both the noise shape and levels. Measurement of noise in low noise devices always presents a challenge, and this is especially true for accurate 1/f noise measurements. High amplification is required for the noise signature to be visible on the measurement equipment. The expected noise floor of the current source was on the order of pA. Across the load of each device, this translated to needing to measure voltages on the order of nV. In order to reduce the dynamic range requirement of the measurement to a practical level, an AC high pass filter with a cutoff frequency close to DC (0.01 - 0.1 Hz) was also required. Typically, this measurement arrangement is formed by combining a high-resolution oscilloscope with a high-gain pre-amplifier. 

A PicoScope 4262\cite{PicoScope:4262} (Pico Technology) was used as the measurement device. It features 16-bits of vertical resolution and an internal pre-amplifier allowing for measurements at 2~mV per division in a 5~MHz bandwidth. To reach the desired measurement range, the pre-amplifier required a gain in excess of 10000. It also required 1/f noise to be lower than the device under test. We have developed our own pre-amplifier to meet these requirements. Details on the pre-amplifier design and performance can be found in the appendix.

The test setup for noise measurement is presented in Fig.~\ref{fig:Test_setup}~a).
The current driver was powered from a Keysight E3630A\cite{Keysight:E3630A} series power supply and controlled through a laptop. The device was connected through an external USB isolator to avoid any potential ground loops that could upset the measurement. For LCD measurements, the output was terminated using a 0.1\% 300~$\Omega$ terminator to convert the current into a voltage that can be amplified by the custom pre-amplifier. For the HCCD the voltage across the $R_F$ resistor can be used as the input to the pre-amplifier, with $R_L$ replaced by a 1~$\Omega$ 50~W dummy load. Measuring across $R_F$ instead of $R_L$ removes the need to use a device capable of measuring a floating voltage. In both cases, the amplified signal was then captured by the oscilloscope.

Due to how the pre-amplifier is constructed, the input capacitor takes around 5-10 minutes to achieve the desired level of low leakage for the reading to stabilise around zero. The oscilloscope was set to DC coupling for each test as the internal AC coupling effectively blocks frequencies below 10 Hz, making it unusable for these tests.

To confirm that the noise floor of the measurement system was below the noise floor of the device under test, the pre-amplifier input was terminated with a 300~$\Omega$ terminator. One hundred seconds of data were captured, at 10~kS/s to ensure that there were enough samples to capture the low-frequency components present in the signal.

\begin{figure}[htb]
	\centering
	\includegraphics[width=0.482\textwidth]{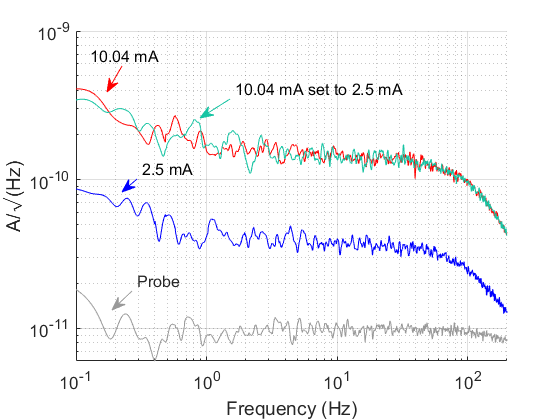}
	\caption{NASD (Noise amplitude spectral density) results in a bandwidth of 0.1 - 200~Hz for the LCD. 10~mA configuration (red), 2.5~mA configuration (blue), 10~mA configuration set to output 2.5~mA (turquoise) and noise floor of the setup terminated with a 300~$\Omega$ resistor (light grey). Channel~3 is the only one shown to improve readability. Other channel results are presented in Table~\ref{tab:noise_tab}. The NASD was obtained using a logarithmic frequency axis power spectral density (LPSD) algorithm \cite{TROBS2006120} made out of 2048 FFT points using a Hann window with amplitude scaling correction applied.
	}
	\label{fig:psd_results}
\end{figure}

After verifying the measurement system, an LCD device was tested. The LCD was populated to have a maximum output current of 10.04~mA and set to output the maximum 10.04~mA on one channel. After the signal was captured the channel was reset back to 0~mA and the process was repeated for the remaining channels. The noise density of a single channel is presented in Fig.~\ref{fig:psd_results}. Key results for all channels are presented in Table.~\ref{tab:noise_tab}.

It can be seen that the signal is not limited by the noise floor of the measurement setup and that the 1/f noise effectively vanishes after 1~Hz, leaving only wideband noise. The average noise floor was calculated at $10\pm5$~Hz, as it is in the flat band between the 1/f noise and the high-frequency roll-off of the device.

To achieve the lowest possible noise it is important to configure the maximum output of each current driver to the lowest possible limit required by the application. To demonstrate this, the previous test was repeated using the same LCD  but with the current set to 1/4 full-scale (2.5~mA). The result is shown in Fig.~\ref{fig:psd_results} (turquoise). The trace overlaps the full-scale output trace, showing there is no improvement in absolute noise performance by programmatically reducing the output current.

To test the scalability of the device, a second LCD was built with the maximum output current set to 2.5~mA. This was achieved by swapping four resistors and a capacitor in the driving stage of each channel. The noise test was re-run and the results are presented in Fig.~\ref{fig:psd_results} (blue) and Table.~\ref{tab:noise_tab}. Compared to the 10~mA LCD, a 4x reduction in maximum current resulted in an approximately 4x reduction in wideband noise. This result shows that the ppb relative scaling has been maintained, however it is expected that for lower output configurations Johnson noise and other noise sources present in the design would start to dominate and would no longer scale with the maximum current output. The optimum peak output current can be selected on a per-device basis based on a balance between absolute noise performance, relative noise performance and minimum step size.

\begin{figure}[htb]
	\centering
	\includegraphics[width=0.482\textwidth]{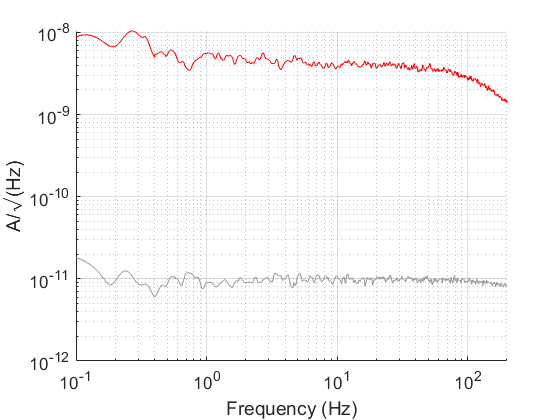}
	\caption{NASD results in a bandwidth of 0.1 - 200~Hz for the HCD. 250~mA configuration (red) and noise floor of the setup terminated with 300~$\Omega$ resistor (light grey). Channel~3 is the only one shown to improve readability. Other channel results are presented in Table~\ref{tab:noise_tab}. The NASD was obtained using logarithmic frequency axis power spectral density (LPSD) algorithm \cite{TROBS2006120} made out of 2048 FFT points using Hann window with amplitude scaling correction applied.}
	\label{fig:psd_results_HCCD}
\end{figure}

The noise test was performed on an HCD that was configured to provide 250~mA full-scale current. The results of this are presented in Fig.~\ref{fig:psd_results_HCCD} and Table~\ref{tab:noise_tab}. Compared to the 10~mA LCD, a 25 times increase in maximum current results in only a minor degradation of relative noise performance.

The noise testing methodology was verified externally using an independent OPM to measure the magnetic noise contribution when the current sources were used to drive the static field coils.

\subsection{Stability test}
For the long term stability test, the oscilloscope and pre-amplifier were replaced with a Keysight B2901A\cite{Keysight:B2901A} precision source meter unit (SMU). The current was measured directly without the use of terminators or dummy loads as the SMU effectively becomes the load. The SMU was expected to have a smaller impact on stability in comparison to the pre-amplifier and oscilloscope making it ideally suited for low cadence long term testing. This feature however prevents it from being used for the bandwidth of interest noise measurements, as its sampling rate of 10 Hz is insufficient.
The test setup for stability measurement is presented in Fig.~\ref{fig:Test_setup}~b).

The stability test was first performed on the LCD configured to provide a full-scale current of 10.04~mA. It was then set to provide about 10\% of its full-scale output (1.004~mA), and monitored for 24 hours. The reason for choosing this output was to maximise the resolution of the SMU which changes its range and loses a digit after 1.1~mA. As shown before in the noise testing section, setting a smaller current without re-configuring the device does not impact its noise performance. The device was set to take 345600 measurements at a sampling frequency of 4~Hz. This frequency was selected to ensure that the SMU analogue to digital converter had enough time to fully settle. This test was repeated for the HCD configured to provide 250~mA and set to output 30\% of its range. Both 1~mA and 100~mA test ranges of the SMU were also tested by shorting the input of the device, to get an estimate of the stability of the measurement system itself.

\begin{figure}[htb]
	\centering
	\includegraphics[width=0.482\textwidth]{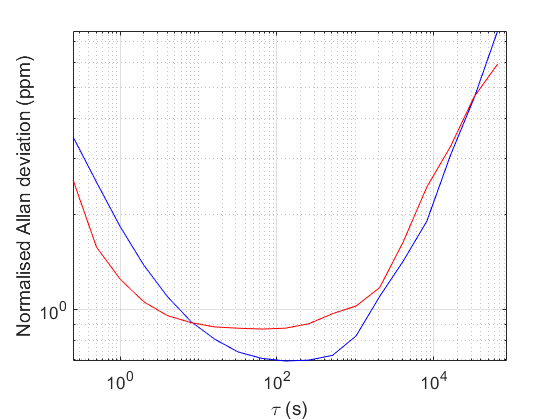}
	\caption{Allan Deviation of both devices, set to 10\% of their full-scale output for 24 hours. LCD (blue) and HCD (red). The first point of the graph shows the ppm stability performance that matches well with the data obtained using the pre-amp and oscilloscope. The improvements from averaging are most notable after around 100 seconds. The device exhibits a long term low frequency drift, but even after 24 hours, it does not drift far enough to compromise its performance.}
	\label{fig:allan_results}
\end{figure}

The stability was estimated using the Allan deviation and later normalised by dividing the result by the measurement range as seen in Fig.~\ref{fig:allan_results}. It can be seen that both devices perform similarly and achieve their best performance after around 100 seconds of averaging, followed by a long term low frequency drift. It is important to note that the ppm stability performance achieved at the start is maintained for at least a working day. The shorted SMU produced better stability for both ranges, indicating that the measurement system was not the limiting factor. It is expected that the long term drift could be mitigated by using components with lower temperature coefficients and zero-drift amplifiers, however doing so could potentially compromise the wideband performance of the device. It is important to state that the long term stability was not a driving requirement, as most of our applications only require a relatively short term stability (less than one hour).

\begin{table*}[t]
	\caption{\label{tab:noise_tab}Noise performance summary of different configurations of the current source}
	\begin{ruledtabular}
		\begin{tabular}{cccc}
Configuration & CH1 & CH2 & CH3 \\

\hline

(LCD) 10~mA, ($10\pm5$~Hz average noise) & 151.1~pA/$\sqrt{\mathrm{Hz}}$ & 147.4~pA/$\sqrt{\mathrm{Hz}}$ & 146.3~pA/$\sqrt{\mathrm{Hz}}$ \\

(LCD) 2.5~mA, ($10\pm5$~Hz average noise) & 37.6~pA/$\sqrt{\mathrm{Hz}}$ & 38.2~pA/$\sqrt{\mathrm{Hz}}$ & 37.7~pA/$\sqrt{\mathrm{Hz}}$ \\

(HCD) 250 mA, ($10\pm5$~Hz average noise) & 4114.2~pA/$\sqrt{\mathrm{Hz}}$ & 4125.2~pA/$\sqrt{\mathrm{Hz}}$ & 4124.1~pA/$\sqrt{\mathrm{Hz}}$ \\

(LCD) 10~mA, ($10\pm5$~Hz relative average noise) & 15.05~ppb/$\sqrt{\mathrm{Hz}}$ & 14.68~ppb/$\sqrt{\mathrm{Hz}}$ & 14.57~ppb/$\sqrt{\mathrm{Hz}}$ \\

(LCD) 2.5~mA, ($10\pm5$~Hz relative average noise) & 15.06~ppb/$\sqrt{\mathrm{Hz}}$ & 15.29~ppb/$\sqrt{\mathrm{Hz}}$ & 15.10~ppb/$\sqrt{\mathrm{Hz}}$ \\

(HCD) 250 mA, ($10\pm5$~Hz relative average noise) & 16.46~ppb/$\sqrt{\mathrm{Hz}}$ & 16.50~ppb/$\sqrt{\mathrm{Hz}}$ & 16.50~ppb/$\sqrt{\mathrm{Hz}}$ \\

		\end{tabular}
	\end{ruledtabular}
\end{table*}

\begin{figure}[htb]
	\centering
	\includegraphics[width=0.482\textwidth]{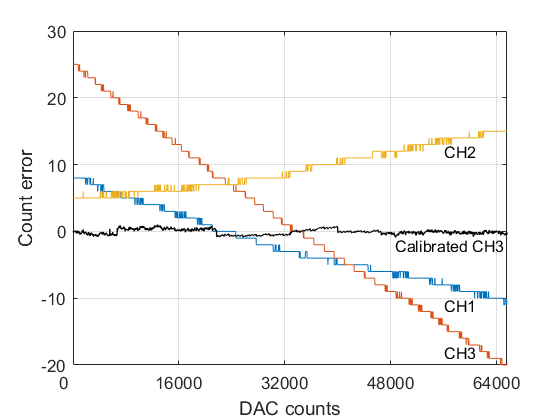}
	\caption{Device count accuracy results, all channels were calibrated but channel~3 is presented for clarity. The result of calibration shows a maximum error of $\pm$1 LSB.}
	\label{fig:calibration_results}
\end{figure}
\subsection{Accuracy test}

A scheme was devised to test the current driver accuracy that used the same test setup as the stability testing outlined in Fig.~\ref{fig:Test_setup}~a). The current source was programmed to go through its 16-bit range in 32 counts steps. This step size was sufficient to show any gain error without the need to test every possible DAC value. The SMU needs at least 200~ms to fully settle making a sweep through a full range possible but impracticable, as it would take approximately 3.5 hours per channel. Sweeping through a single channel with a step size of 32 (for a total of 2048 steps) takes approximately 7 minutes, totalling 21 minutes for all of the channels.

The data was then compared to the theoretical accuracy given by Eq.~(\ref{eq:1}) and the difference was converted into DAC counts. The results are shown in Fig.~\ref{fig:calibration_results}.

The results show that the main source of error is the gain error (demonstrated by the non-zero gradient of the line) followed by the offset error. A lookup table (LUT) based compensation method was implemented on the Arduino. The LUT was populated using the error count derived from the calibration routine, with one correction value for each 32 count block. Testing showed that the calibration routine increased the accuracy to within $\pm$1 least significant bit (LSB) ($\pm$15.3 ppm), as shown for channel~3 in Fig.~\ref{fig:calibration_results}. It is important to note that the baseline accuracy exhibited at worst a $\pm$25 count error with no alteration to the circuit. This result is notable when considering that no matching of components was performed. This shows that the device can be used without an explicit need for calibration, as the percentage error is less than 0.04\% which is negligible for our application. Note that 2048 steps were not actually needed to resolve the gain error and that potentially the data could be compressed into bigger chunks which would speed up the calibration process.

\begin{figure*}[!t]
	\centering
	\includegraphics[width=1\textwidth]{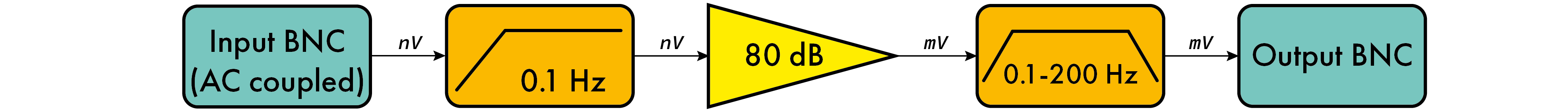}
	\caption{Simplified architecture of the pre-amplifier. The input signal is buffered and stripped of its DC component with a 0.1 Hz HPF. It is then followed by a 80 dB gain block that is further filtered by a 0.1 - 200 Hz bandpass filter. The resulting signal is available on the output BNC for connection to an oscilloscope.}
	\label{fig:probe_setup}
\end{figure*}

\begin{figure}[!t]
	\centering
	\includegraphics[width=0.482\textwidth]{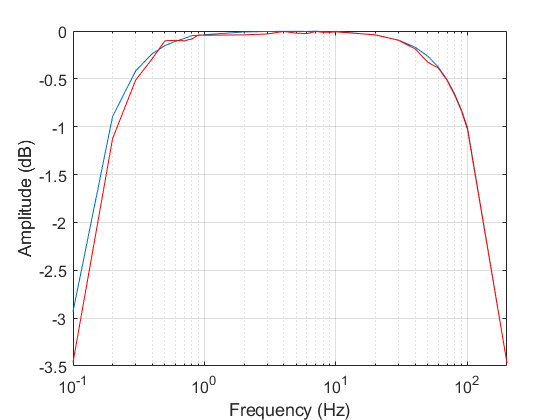}
	\caption{Experimental (red) and simulated (blue), normalised pre-amplifier frequency response (0.1 - 200 Hz). The frequency response exhibits 50 mdB flatness. The simulated results match well with the experimental ones. The biggest difference being the corner frequency at the low end which is different due to the inherent tolerance of the electrolytic input capacitor. This however does not impact the overall bandwidth of the pre-amplifier.}
	\label{fig:probe_bw}
\end{figure}

\section{Conclusions}

This paper described the design procedure and testing of two ultra-low noise current drivers. The LCD is capable of producing currents up to \textasciitilde$\pm$~50mA and the HCD is capable of providing \textasciitilde$\pm$~250~mA. The devices complement each other by overlapping the gap between current ranges, allowing them to be used in a variety of applications. Both devices offer digital control of current on three independent axis with 16-bit resolution. They feature a common digital/signal conditioning chain allowing for user customisation in terms of bandwidth, noise or stability. This was demonstrated by changing the maximum output current capability which allowed the device to change its dynamic range. When used to drive coils, having a selection of devices to choose from helps match the driver with a given coil geometry. An Arduino based firmware makes the device control extendable to non-expert users and allows users to extend the functionality to their particular application such as generation of arbitrary waveforms.
In addition, both devices are fully open source, with documentation and design files being readily available on GitHub \cite{github_current_source} (link in appendix).

Noise tests showed that a 10.04~mA LCD achieved 146.3~pA/$\sqrt{\mathrm{Hz}}$ at 10~Hz, demonstrating a relative noise of 14.57~ppb/$\sqrt{\mathrm{Hz}}$. A 250~mA HCD achieved 4114~pA/$\sqrt{\mathrm{Hz}}$ at 10~Hz, demonstrating a relative noise of 16.46~ppb/$\sqrt{\mathrm{Hz}}$. Both devices feature a narrow 1/f region with a corner frequency of approximately 1~Hz, making them well suited for precision magnetic field generation in OPMs.

The stability of both devices was found to be in the ppm range. Due to the fact that the only difference between LCD and HCD is their output stage, the stability performance is very similar. This performance is more than enough for coil control in most OPM applications which primarily rely on short term stability.

Accuracy tests showed that the device does not have to be calibrated to achieve satisfactory performance, presenting an out of the box solution. For applications demanding very high accuracy it was shown that calibration is possible, unlocking the full potential of the device with monotonic accuracy of $\pm$1 LSB over its full range.

\begin{acknowledgments}
	The work was supported by the UK Engineering and Physical Sciences Research Council under grant number EP/T001046/1.
\end{acknowledgments}

\section*{Data Availability Statement}
All design files, datasets and analysis scripts are available on the project GitHub\cite{github_current_source}. Release v1.1.1 was used for this paper. An archived version is available via Zenodo\cite{iain_chalmers_2022_6833316}.

\FloatBarrier
\appendix*
\section{High gain pre-amplifier design}
The pre-amplifier was based on a design developed for measuring the noise contribution of an LTC6655 reference \cite{linear}. It was modified and modernised to replace the high gain stage that contained difficult to procure components, such as thermally-lagged JFET pairs and wet tantalum capacitors. 

The simplified architecture of the pre-amplifier is presented in Fig.~\ref{fig:probe_setup}. The design is centred around ADA4523-1\cite{ADI:ADA4523-1} (Analog Devices) chopper amplifiers, which form the gain stage. These exhibit very low noise at low frequencies. This is achieved by heterodyning  1/f noise on top of a chopping frequency of 330 kHz, which is located away from the 0.1 to 160 Hz bandwidth of interest. The signal first goes through an optional buffer that presents the load with a high impedance input. The DC component is then removed using a hand-selected, sub-10-nA leakage current electrolytic capacitor to provide a high pass filter (HPF) at 0.1 Hz. The HPF stage is very important, as one cannot rely on the oscilloscope AC-coupled input because of its high cut-off frequency (>10~Hz) which would obscure the 1/f noise contribution that needs to be measured. The gain stage consists of four ADA4523-1 amplifiers connected in parallel to reduce the voltage noise by a factor of 2 and provide a gain of 10000 (80 dB). The signal is later conditioned by a second-order Sallen-Key LPF with a Bessel response to reduce ringing at the output \cite{2005323}. The final stage is a passive HPF to further improve the 0.1 Hz response. The device is interfaced with BNC connectors on each side and powered by two 6LR61 batteries. It provides a gain of 80 dB in a bandwidth of 0.1 - 200 Hz as seen in Fig.~\ref{fig:probe_bw}. The pre-amplifier design is open source and available on GitHub\cite{github_probe}.

\FloatBarrier
\section*{References}
\bibliography{current_source_paper}

\end{document}